\documentclass[%
preprint,
 amsmath,superscriptaddress,amssymb,aps,
]{revtex4-2}
\usepackage{xcolor}
\usepackage{booktabs}
\usepackage[section]{placeins}
\usepackage{graphicx}
\usepackage{dcolumn}
\usepackage{bm}

\begin{document}


\title{Thermal spin current generation in the multifunctional ferrimagnet Ga$_{0.6}$Fe$_{1.4}$O$_{3}$}

\author{Alberto Anadón}
 \affiliation{Institut Jean Lamour, Université de Lorraine CNRS UMR 7198, Nancy, France}
\email{alberto.anadon@univ-lorraine.fr, juan-carlos.rojas-sanchez@univ-lorraine.fr}
\author{Elodie Martin}
\affiliation{Institut Jean Lamour, Université de Lorraine CNRS UMR 7198, Nancy, France}
\author{Suvidyakumar  Homkar}
\affiliation{Université de Strasbourg, CNRS, IPCMS, UMR 7504, F-67000 Strasbourg, France}
\author{Benjamin  Meunier}
\affiliation{Université de Strasbourg, CNRS, IPCMS, UMR 7504, F-67000 Strasbourg, France}
\author{Maxime Verges}
\affiliation{Institut Jean Lamour, Université de Lorraine CNRS UMR 7198, Nancy, France}
\author{Heloise  Damas}
\affiliation{Institut Jean Lamour, Université de Lorraine CNRS UMR 7198, Nancy, France}
\author{Junior Alegre}
\affiliation{Institut Jean Lamour, Université de Lorraine CNRS UMR 7198, Nancy, France}
\author{Christophe   Lefevre}
\affiliation{Université de Strasbourg, CNRS, IPCMS, UMR 7504, F-67000 Strasbourg, France}
\author{Francois   Roulland}
\affiliation{Université de Strasbourg, CNRS, IPCMS, UMR 7504, F-67000 Strasbourg, France}
\author{Carsten Dubs}
\affiliation{INNOVENT e.V. Technologieentwicklung, Jena, Germany}
\author{Morris Lindner}
\affiliation{INNOVENT e.V. Technologieentwicklung, Jena, Germany}
\author{Ludovic Pasquier}
\affiliation{Institut Jean Lamour, Université de Lorraine CNRS UMR 7198, Nancy, France}
\author{Olivier Copie}
\affiliation{Institut Jean Lamour, Université de Lorraine CNRS UMR 7198, Nancy, France}
\author{Karine Dumesnil}
\affiliation{Institut Jean Lamour, Université de Lorraine CNRS UMR 7198, Nancy, France}
\author{Rafael Ramos}
\affiliation{Centro de Investigación en Química Biolóxica e Materiais Moleculares (CIQUS), Departamento de Química-Física, Universidade de Santiago de Compostela, 15782, Spain.}
\author{Daniele Preziosi}
\affiliation{Université de Strasbourg, CNRS, IPCMS, UMR 7504, F-67000 Strasbourg, France}
\author{Sébastien Petit-Watelot}
\affiliation{Institut Jean Lamour, Université de Lorraine CNRS UMR 7198, Nancy, France}
\author{Nathalie Viart}
\affiliation{Université de Strasbourg, CNRS, IPCMS, UMR 7504, F-67000 Strasbourg, France}
\author{Juan-Carlos Rojas-Sánchez}
\affiliation{Institut Jean Lamour, Université de Lorraine CNRS UMR 7198, Nancy, France}
\date{\today}

\begin{abstract}
In recent years, multifunctional materials have attracted increasing interest for magnetic memories and energy harvesting applications. Magnetic insulating materials are of special interest for this purpose, since they allow the design of more efficient devices due to the lower Joule heat losses. In this context, Ga$_{0.6}$Fe$_{1.4}$O$_3$ (GFO) is a good candidate for spintronics applications, since it can exhibit multiferroicity  and presents a spin Hall magnetoresistance similar to the one observed in a yttrium iron garnet (YIG)/Pt bilayer. Here, we explore GFO utilizing thermo-spin measurements in an on-chip approach. By carefully considering the geometry of our thermo-spin devices we are able to quantify the spin Seebeck effect and the spin current generation in a GFO/Pt bilayer, obtaining a value comparable to that of YIG/Pt. This further confirms the promises of an efficient spin current generation with the possibility of an electric-field manipulation of the magnetic properties of the system in an insulating ferrimagnetic material.\end{abstract}

\maketitle

\section{\label{sec:intro}Introduction}

The search for multifunctional materials is nowadays a hot topic in spintronics\cite{eerenstein2006multiferroic,scott2007multiferroic}. Currently, functional devices are typically made of a bilayer composed of a non-magnetic material with large spin-orbit coupling (NM) and a ferromagnetic material (FM). These types of devices allow functionalities such as the manipulation of the FM magnetization by the spin Hall effect (SHE)\cite{avci2017current,garello2014ultrafast,anadon2020spin} in the NM or energy harvesting by employing its inverse counterpart, the inverse spin Hall effect\cite{ramos2016thermoelectric}. Insulating magnetic materials (FMI) are preferred for this purpose to pave the way towards low dissipation spintronics devices\cite{avci2017current}. Additional functionalities like the possibility of the electric field control of the magnetic properties of such systems could be given to these heterostructures by the introduction of multifunctional magnetic materials, opening the possibility of designing more efficient and versatile devices\cite{Manipatruni2019,ramesh2009emerging}. 

In spin Seebeck experiments\cite{uchida2014longitudinal,anadon2021engineering,uchida2010spin,Uchida2016} a thermal gradient is typically applied in the out-of-plane direction of the magnetic thin film, generating a spin current flowing alongside this direction. In insulating ferromagnetic materials, this spin current is carried by spin collective excitations, also called magnons, and can be injected into an adjacent layer such as Pt in the case of this study, and is then converted into a charge current via the inverse spin Hall effect (ISHE)\cite{hoffmann2013spin,sinova2015spin,Rezende2014}. This conversion occurs through the spin-orbit interaction of conduction electrons, which can be strong in heavy metals like Pt and is given by\cite{kikkawa2021observation}:

\begin{equation}
    \mathbf{E}_{ISHE}=\frac{2e}{\hbar}\rho\cdot\theta_{SH}\mathbf{J}_s\times\mathbf{\sigma},
\end{equation}
where $\mathbf{E}_{ISHE}$ is the electric field produced by the ISHE, $e$ and $\hbar$ are the electron charge and reduced Plank constant, $\rho$ is the resistivity of the Pt layer, $\theta_{SH}$ is the spin Hall angle, $\mathbf{J}_s$ is the spin current injected into Pt and $\sigma$ its spin polarization.

Until now, yttrium iron garnet (YIG) has been the cornerstone material in thermo-spin phenomena due to its insulating nature and its unique magnetic properties such as low damping and coercive field. Here, we have studied the thermo-spin current generation in bilayers composed of Pt and the multifunctional magnetoelectric oxide Ga$_{0.6}$Fe$_{1.4}$O$_3$ (GFO)\cite{roy2014structure,stoeffler2012first}.
Engineering of thermo-spin devices with properly chosen dimensions allowed an accurate determination of the thermo-spin voltages necessary to calculate the spin Seebeck coefficient, and therefore granting a comparison with other systems. Indeed, the spin Seebeck effect (SSE) as well the spin Hall magnetoresistance \cite{Althammer2013,Nakayama2013,homkar2021spin} in GFO system are largely comparable with YIG-based ones. Furthemore, to corroborate our experimental finding we resorted to finite element simulations of the thermal profile to obtain an accurate heat flux in both GFO and YIG layers.

\section{\label{sec:methods}Methods}
\subsection{Growth and structural characterization}

\begin{figure*}
\includegraphics[width=0.9\linewidth]{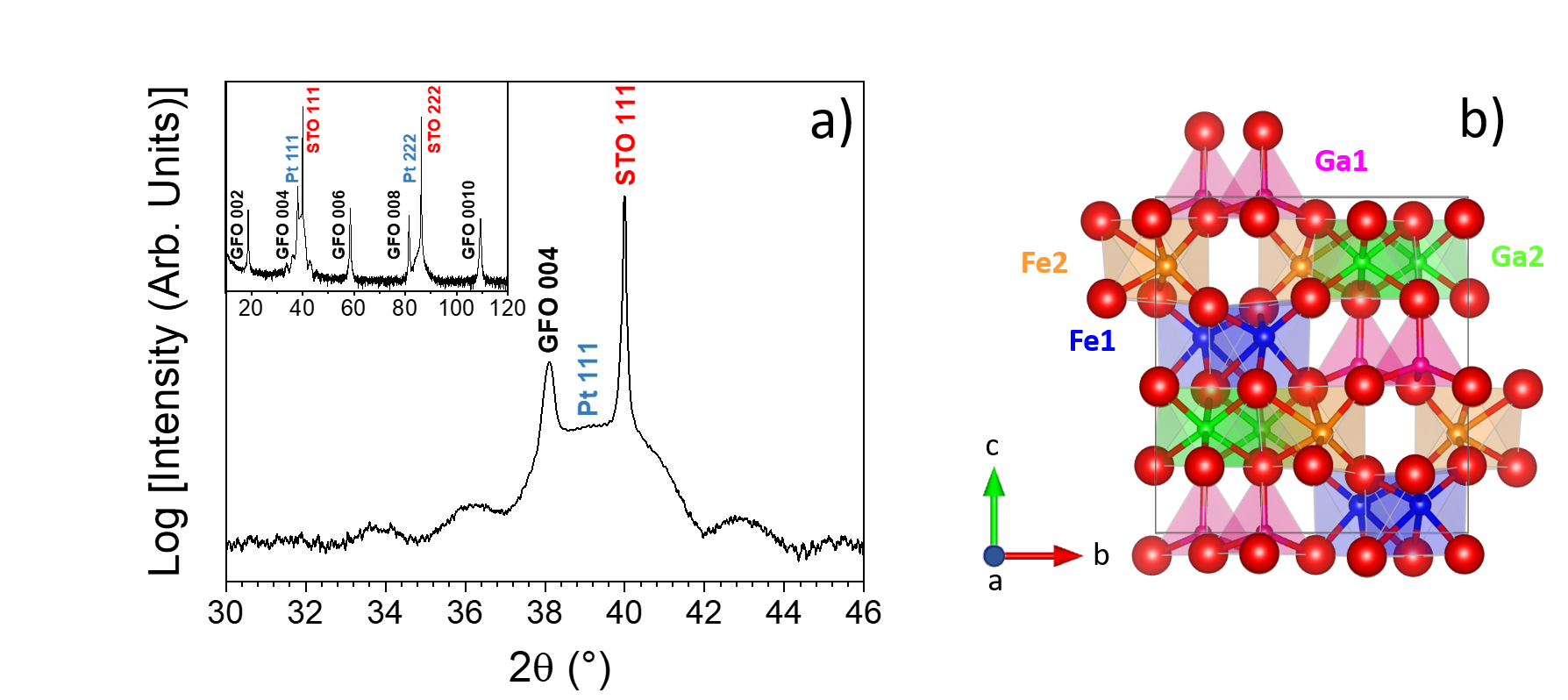}
\caption{\label{figstructure} \textbf{Structural characterization.} a) X-ray diffraction $\theta-2\theta$ scan showing the Ga$_{0.6}$Fe$_{1.4}$O$_3$ (004) peak, the Pt (111) peak and the associated Laue oscillations, indicating high crystalline coherence in the film and b) crystal structure of Ga$_{0.6}$Fe$_{1.4}$O$_3$. In the crystallographic structure of GFO there are four different cationic sites that can be occupied by the Ga$^{3+}$ and Fe$^{3+}$ cations, named Ga1, Fe1, Ga2, and Fe2. Ga1 is a tetrahedral site, and the Fe1, Ga2, and Fe2 are non-equivalent octahedral sites.}
\label{figcharacterization}
\end{figure*}

GFO films were prepared by pulsed laser deposition (PLD) on SrTiO$_3$(111)  substrates (Furuuchi Chemical Corporation, Japan, with root mean square roughness lower than 0.15 nm) maintained at 900\textdegree C. The KrF excimer laser ($\lambda$ = 248 nm) was operated with a fluence of 4 J/cm$^2$ \cite{homkar2021spin} and a repetition rate of 2 Hz. The growth was done from a stoichiometric GFO target in an atmosphere of 0.1 mBar of O$_2$. The YIG film was grown by liquid phase epitaxy on 3-inch (111)-oriented gadolinium gallium garnet (GGG) substrate from PbO-B$_2$O$_3$-based high-temperature solution (HTL) at about 800\textdegree C using a standard dipping technique \cite{giess1975liquid,dubs2017sub}. During the deposition time of 90 seconds, the substrate was rotated in the solution at 33 rpm and then pulled out. Subsequently, the solution residues were spun off from the sample surface above the HTL and the sample was pulled out of the hot heating zone. Platinum layers for the GFO-based samples were deposited by PLD in situ in order to avoid any surface contamination. The deposition was carried out at room temperature in order to avoid any metal/oxide interdiffusion, under vacuum (base pressure of 2 x 10$^{-8}$ mbar) and with a deposition rate of 0.06 nm/s, as in \cite{homkar2021spin}.

Structural characterization of the samples was done by X-ray diffraction $\theta-2\theta$ scans using a Rigaku Smart Lab diffractometer equipped with a rotating anode (9 kW) and monochromated copper radiation (1.54056 \AA). The $\theta-2\theta$ scan shown in figure \ref{figcharacterization} indicates that the system grows following the (SrTiO$_3$) STO(111)//GFO(001)/Pt(111) structure, i.e., with the GFO film oriented  along the [001] direction. The presence of Laue oscillations for the Pt (111) reflection is an indication of a smooth Pt/GFO interface.

\begin{figure*}
\includegraphics[width=0.8\linewidth]{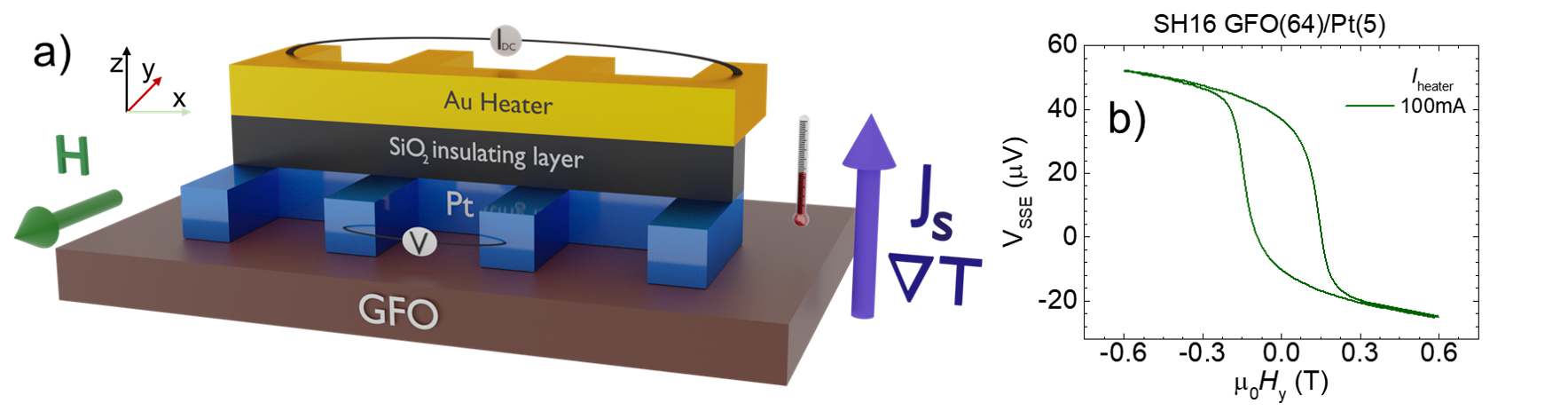}
\caption{\label{figschemeSSE} \textbf{Thermo-spin measurements.} a) Longitudinal Spin Seebeck effect configuration. b) Transversal voltage as a function of the in-plane field in a thermo-spin measurement with a heater current of 100 mA.}
\end{figure*}

The thermo-spin devices were made using conventional UV lithography. We have chosen the following stacking to perform our main experiments: STO//GFO(64 nm)/Pt(5 nm) and GGG//YIG(140 nm)/Pt(5 nm). The Pt thin film is first patterned by ion milling. Then, an insulating SiO$_2$ layer with a thickness of 75 nm is grown by RF sputtering using a Si target and Ar$^+$ and O$^{2-}$ plasma. In a third and last step the Au heater (Ti(10 nm)/Au(150 nm)) is evaporated using a conventional evaporator. Both the heater and the sample are patterned with four pads to measure their resistance using four probes for more precise estimation. The dimensions of the active part of the heater and the sample are 330$\times$10 $\mu$m$^2$ and 270$\times$10 $\mu$m$^2$ respectively. 

\subsection{Thermo-spin measurements and estimation of heat transport parameters}
Thermo-spin measurements are carried out in these devices using an electromagnet to apply an external in-plane magnetic field (H) as shown in figure \ref{figschemeSSE} (a). A DC current is passed through the heater and after 5 minutes of stabilization, the resistances of the sample and the heater are monitored using I-V measurements to avoid spurious contributions from thermal voltages. The thermo-spin voltage is monitored using a Keithley 2182a nano voltmeter. 

Numerical simulations based on finite element method have been performed by COMSOL multiphysics, coupling the Electric Currents and Heat Transfer modules, in order to quantify thermal gradients in GFO on STO substrate and in YIG on GGG substrate (see supporting information). The cross-plane thermal conductivity of the GFO thin film was measured using the 3$\omega$ method. In this method a thin metal resistor simultaneously serves as a heater and a thermometer, it has been previously employed to determine the thermal conductivity of bulk and thin film materials, details of the measurements can be found elsewhere \cite{cahill1990thermal,EricThermal,PilarThermal}. For these measurements a Pt resistor (100 nm thick, 10 um width, 1 mm length with 10 nm of Cr for adhesion) is deposited on the bare GFO/STO heterostructure and the 3$\omega$ voltage response to an AC current with frequency is measured. The thermal parameters of the other layers are obtained from literature\cite{PhysRevB.92.094406,boona2014magnon,suemune1965thermal,bussmann2019simple,de1996high,wang2008thermal,samoshkin2020heat} and detailed in the supporting information.

\section{\label{sec:results}Results}

\subsection{Thermo-spin voltage in GFO/Pt and YIG/Pt}
\begin{figure*}
\includegraphics[width=\linewidth]{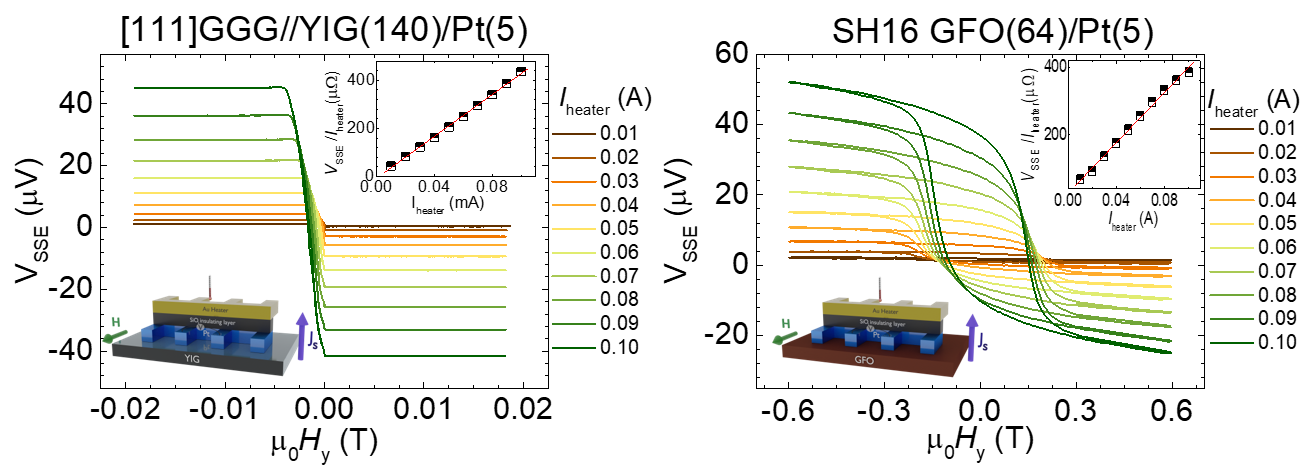}
\caption{\label{fig4} \textbf{Spin Seebeck effect in GFO and YIG.} A comparable spin Seebeck voltage is obtained in the GFO/Pt and in the YIG/Pt system varying the heater power. The insets show the linear behaviour of the thermo-spin voltage with the heater power.}
\end{figure*}

We have performed thermo-spin measurements in a GFO/Pt and YIG/Pt systems by observing the thermally induced voltage upon sweeping H as shown in figure \ref{figschemeSSE}(a) for GFO, obtaining a typical hysteresis loop-like curve that follows the magnetization of the GFO thin film\cite{homkar2021spin} (figure \ref{figschemeSSE}(b)) for a heater current of 100 mA. We have performed these measurements for H applied at different directions in YIG and GFO and observed an isotropic behaviour of the SSE voltage.

By monitoring the magnitude of the thermo-spin voltage at saturation for different heater powers we can observe that the voltage difference between the saturation at positive and negative fields scales linearly with the power applied to the heater, as expected by the origin of the spin current generated in GFO. This is shown in the insets of figure \ref{fig4} for both GFO and YIG
systems. We also observe here that the order of magnitude of the SSE voltage is similar in both systems, although in the GFO the offset voltage of the loop and the coercive field are significantly larger than in YIG. While the elucidation of the origin of such offset voltage is out of the scope of this paper, we argue that it might be related to a persistent pyroelectric current due to the polar nature of the GFO\cite{GFOferroelectric}. 

\subsection{Quantitative comparison of the spin Seebeck effect in YIG and GFO}

 The determination of the spin Seebeck coefficient is normally subjected to large uncertainty due to the low reproducibility of the experimental conditions. Typically, in literature, the SSE coefficient is defined as the ratio between the SSE voltage and the thermal gradient in the FM thin film normalized by the sample resistance\cite{Rezende2018}. 
  \begin{equation}
     S_{SSE}^{\nabla T}=\frac{V^{ISHE}}{\nabla T\cdot R_{Pt}\cdot l},
     \label{sseq1}
 \end{equation}
 where V$^{ISHE}$ is the thermo-spin voltage difference between positive and negative saturation fields divided by two, $\nabla$T is the thermal gradient through the FMI, R$_{Pt}$ is the 4-probe resistance in the Pt using I-V measurements and \emph{l} is the distance between the voltage contacts.
 
 Two steps can be taken to overcome the issue of low reproducibility in SSE measurements: first, the heat flux in the system can be considered instead of an estimation of the thermal gradient\cite{Sola2017}, and second, it is possible to maximize the reproducibility of the thermal conditions by using on-chip devices for a consistent thermal contact and dimensions of the system\cite{wu2015spin,PhysRevB.103.L020401,doi:10.1126/sciadv.abg1669}. 
 
Following both of these steps, we propose to define the SSE coefficient considering the heat flux instead of the thermal gradient in the FMI (S$_{SSE}^q$) as follows:
 \begin{equation}
     S_{SSE}^{\phi_q}=\frac{V^{ISHE}}{\phi_q\cdot R_{Pt}},
     \label{sseq}
 \end{equation}
where $\phi_q$ is the heat flux through the FMI.
 
 \begin{figure*}
\includegraphics[width=\linewidth]{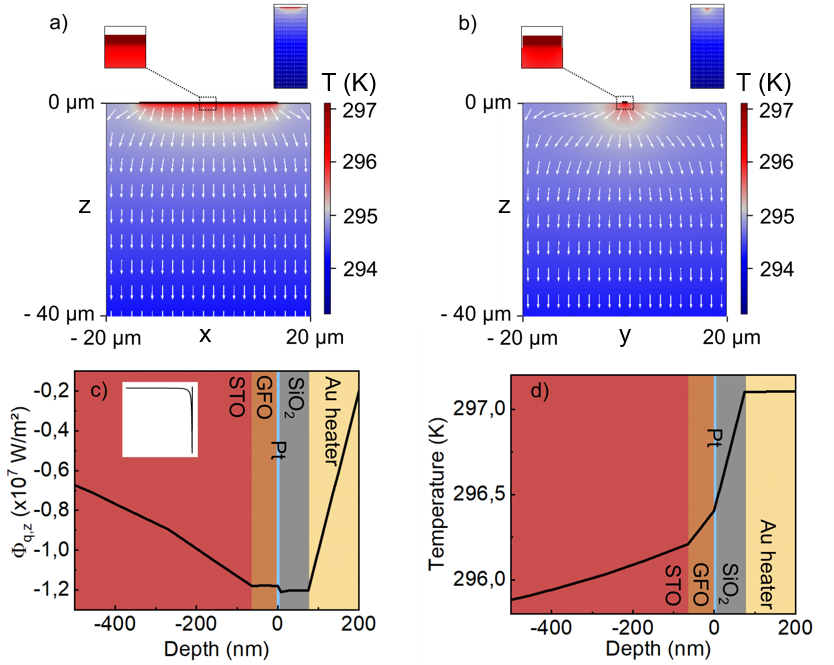}
\caption{\label{fig3} \textbf{Thermal simulation profile for the STO//GFO/Pt thermo-spin device.} Calculated  local temperature within the device that results from heating: a) (X0Z) cut plane and b) (0YZ) cut plane. The insets show the same distribution near the device (top left) and at an expanded range through the whole system (top right). c) z-component of heat flux in the out-of plane direction at the center of the device as calculated by the numerical method. The inset extends the range in depth including the substrate, showing that the heat flux close to the bottom of the substrate vanishes. d) Temperature  in the out-of-plane direction at the center of the device as calculated by the numerical method.}
\end{figure*}


 To  reliably compare the thermo-spin voltage in the GFO/Pt and YIG/Pt systems we compute the S$_{SSE}^{\phi_q}$ from equation \ref{sseq} and obtain a comparable value for both systems as depicted in table \ref{tableSSC}. We obtain a S$_{SSE}^{\phi_q}$= 2.9$\pm$0.3 $\frac{fV\cdot m^2}{W\cdot \Omega}$ for GFO, whereas for YIG the value is slighty larger, 3.6$\pm$0.4. This suggests that the efficiency of the spin current generation in GFO is similar to that of YIG, opening new possibilities in spin caloritronics using insulating magnetic materials with predicted tunable magnetic properties by electric fields.
 
 Using the thermal conductivity for both FMI we can also estimate the thermal gradient in the FMI layers considering Fourier's law and a one-dimensional thermal flux ($\mathbf{\phi_q}=-\kappa\nabla T$) to compare with other studies in literature as shown in table \ref{tableSSC}. We observe that the temperature difference between the upper and lower boundaries of the FMI in both systems are similar in magnitude under these considerations. To assess possible deviations from this simplified model due to the geometry of the device and possible thermal losses, we have performed a finite element simulation using COMSOL. In this simulation, we introduce the geometry of the system and use a combination of the electric currents and heat transfer modules and we obtain a temperature difference in the FMI layer slightly smaller but comparable to the one calculated by Fourier's law for both systems. Figures \ref{fig3}(a) and (b) show respectively the local temperature  distribution in the device in both X0Z and 0YZ cut planes. They show that the temperature gradient direction is mostly out of plane within the device, as expected. The heat flux through the GFO layer is almost constant as shown in figure \ref{fig3}(c) indicating that the thermal losses are not relevant and the direction of the thermal gradient is almost completely out of plane in the GFO. Figure \ref{fig3}(d) shows the simulated thermal profile in the whole device. 
 
 \begin{table}[ht]
\centering
\begin{tabular}{@{}cccccccc@{}}
\toprule
       & $\rho_{Pt}$ ($\mu\Omega$cm)&  \textbf{S$_{SSE}^{\phi_q}$ ($\frac{fV\cdot m^2}{W\cdot \Omega}$) } & $\kappa_{FMI}$ (W/m$\cdot$K) & $\nabla T_{FMI}^{Fourier}$ (K)& $\nabla T_{FMI}^{COMSOL}$ (K) &  S$_{SSE}^{\nabla T}$ ($\frac{pV}{K\cdot \Omega}$)\\ \midrule
GFO/Pt & 27.5 &            \textbf{2.9$\pm$0.3}    &              4 $\pm$ 1        &     0.251      &       0.191                                     &                    90   $\pm$  10        \\
YIG/Pt & 25.3 &     \textbf{3.6$\pm$0.4}     &             8.5               &          0.252     &           0.190                                 &                    160   $\pm$ 20        \\ \bottomrule
\end{tabular}
\caption{\label{tableSSC}\textbf{Thermo-spin and thermal transport parameters for YIG and GFO at room temperature}. Electrical resistivity, heat flux spin Seebeck coefficient, thermal conductivity of the FMI, estimated temperature difference between the upper and lower boundaries of the FMI layer for a heater current of 100 mA using Fourier's law and COMSOL simulations and thermal gradient spin Seebeck coefficient. The thickness of the FMI are 64 and 140 nm for the GFO and YIG films respectively.}
\end{table}

   In the case of GFO, we have estimated the thermal conductivity using the 3$\omega$ method to be $\kappa_{GFO}=4 \pm 1$ W/mK, significantly smaller than that of YIG\cite{PhysRevB.92.094406}. The rest of the values considered have been extracted from literature \cite{PhysRevB.92.094406,boona2014magnon,suemune1965thermal,bussmann2019simple,bussmann2019simple,de1996high,wang2008thermal,samoshkin2020heat}. Following these estimations, the value of the SSE considering the thermal gradient in the FMI can be recovered to compare with other studies. The calculated values of S$_{SSE}^{\nabla T}$ are 90$\pm$10 and 160$\pm$20 $\frac{pV}{K\cdot \Omega}$ respectively. The rest of the parameters to obtain them are shown in table \ref{tableSSC}. We obtain a value of S$_{SSE}^{\nabla T}$ within the same order of magnitude and a similar value of S$_{SSE}^{\phi_q}$ for both materials, showing that for both methods of estimating the spin current generation point towards the interest in using GFO as a promising functional material in insulating spintronics.

\section{\label{sec:Conclusions}Conclusions}
In summary, we have explored the thermo-spin properties of the multi-functional material Ga$_{0.6}$Fe$_{1.4}$O$_3$ in the form of thin film for energy harvesting and thermal management applications. This material provides additional functionality in terms of ferroelectricity while maintaining the electrically insulating behaviour compared to other materials such as yttrium iron garnet. By using an on-chip approach to increase reproducibility and carefully considering the heat flux and thermal gradients in both systems, we find that the spin current generation by thermal excitation is comparable to the one of yttrium iron garnet. This observation supports the promises of an efficient spin current generation with the possibility of an electric-field manipulation of the magnetic properties of the system in a new insulating ferrimagnetic material. Our results show that this material can be exploited in spintronics and spin caloritronics applications.

\begin{acknowledgments}
This work was funded by the French National Research Agency (ANR) through the ANR-18-CECE24-0008-01 ‘ANR MISSION’ and the No. ANR-19-CE24-0016-01 ‘Toptronic ANR’. S.H. acknowledges the Interdisciplinary Thematic Institute QMat, as part of the ITI 2021 2028 program of the University of Strasbourg, CNRS and Inserm, supported by IdEx Unistra (ANR 10 IDEX 0002), SFRI STRAT’US project (ANR 20 SFRI 0012), ANR-11-LABX-0058 NIE, and ANR-17-EURE-0024 under the framework of the French Investments for the Future Program. M. L. acknowledges the funding by the German Bundesministerium für Wirtschaft und Energie (BMWi) - 49MF180119. R.R. acknowledges support from the European Commission through the project 734187-SPICOLOST (H2020-MSCA-RISE-2016), the European Union’s Horizon 2020 research and innovation programme through the MSCA grant agreement SPEC-894006, Grant RYC 2019-026915-I funded by the MCIN/AEI/10.13039/501100011033 and by “ESF investing in your future”, the Xunta de Galicia (ED431B 2021/013, Centro Singular de Investigación de Galicia Accreditation 2019-2022, ED431G 2019/03) and the European Union (European Regional Development Fund - ERDF). We also thank B. Wenzel, R. Meyer, M. Reich and O. Surzhenko (INNOVENT) for their support.
\end{acknowledgments}
\nocite{*}

\bibliography{apssamp}

\clearpage
\section*{Supporting information}
\subsection{Methods}
\setcounter{figure}{0}
\renewcommand{\figurename}{Fig.}
\renewcommand{\thefigure}{S\arabic{figure}}
\setcounter{table}{0}
\renewcommand{\tablename}{Table}
\renewcommand{\thetable}{S\arabic{table}}
\begin{table}[ht]
\begin{tabular}{@{}lllll@{}}
\toprule
     & t (nm)  & $\kappa$(W/m$\cdot$K)    & Cp (J/Kg$\cdot$K)   & $\rho$(Kg/m$^3$) \\ \midrule
GFO  & 64      & 4$ \pm $1   (this paper)     & 680*           & 5 530     \\
YIG  & 140     & 8.5\cite{PhysRevB.92.094406}  & 600\cite{boona2014magnon}   & 5 170     \\
STO  & 100000 & 12\cite{suemune1965thermal,bussmann2019simple} & 437\cite{bussmann2019simple,de1996high} & 4 891     \\
GGG  & 100000 & 7.5\cite{wang2008thermal}  & 380\cite{samoshkin2020heat}   & 7 080     \\
SiO2 & 75      & 1.3          & 680           & 2 270     \\
Pt   & 5       & 69.1         & 130           & 21 450    \\
Au   & 150     & 310          & 130           & 19 300    \\
Air  & 225     & 0.0262       & 1 003         & 1.225     \\ \bottomrule
\end{tabular}
\caption{\label{tableparameters}Parameters for the thermal simulations using COMSOL.* Specific heat capacity (Cp) of GFO is not available in literature to our knowledge, so the Cp of SiO$_2$ was used.}
\end{table}

Thermal gradients were calculated in GFO on STO substrate and in YIG on GGG substrate by finite element method (FEM) simulations in COMSOL Multiphysics software by coupling Electric Currents and Heat transfer modules. An electric current density j$_c$= 6$\cdot$10 A/m$^2$ flows into the Au wire while temperature is fixed at 293.15 K on the bottom of the substrate. We used thermal insulation condition on all the external boundaries (except the one whose temperature is fixed at room temperature) and thin layer condition for the Pt layer. We considered a Pt(5 nm) / SiO2(75 nm) / Au(150 nm) wire (27 x 1 um²) on either GGG (100 $\mu$m) / YIG (140 nm) or STO (100 $\mu$m) / GFO (64 nm). All the parameters of interest for each material are indicated in table \ref{tableparameters}.
\subsection{Finite element simulation of the thermal profile in YIG}
\begin{figure*}
\includegraphics[width=\linewidth]{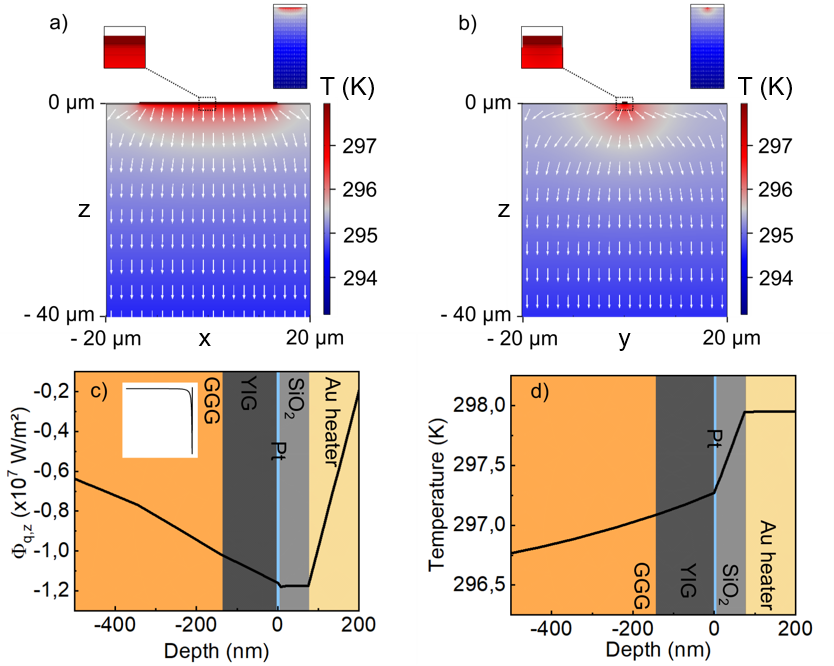}
\caption{\label{figYIG} \textbf{Thermal simulation profile for the GGG//YIG/Pt thermo-spin device.} Calculated  local temperature within the device that results from heating with a) (X0Z) cut plane and b) (0YZ) cut plane. The insets show the same distribution near the device (top left) and at an expanded range through the whole system (top right). c) z-component of heat flux in the out-of plane direction at the center of the device as calculated by the numerical method. The inset presents the same information in the stack including the substrate. d) Temperature  in the out-of-plane direction at the center of the device as calculated by the numerical method.}
\end{figure*}

A finite element simulation was also carried out using the parameters of the YIG/Pt sample as shown in table \ref{tableparameters}. The resulting thermal and heat flux profiles are similar to the ones in GFO and can be observed in figure \ref{figYIG}. 

\end{document}